\documentclass[twocolumn,tighten,numberedappendix]{aastex63}
\usepackage{amsmath}
\usepackage{ulem}
\usepackage{hyperref}
\setcitestyle{notesep={ }}

\graphicspath{{./}{figures/}}

\definecolor{ao(english)}{rgb}{0.0, 0.5, 0.0}

\newcommand{\emin}{\epsilon_\mathrm{min}}
\newcommand{\emax}{\epsilon_\mathrm{max}}
\newcommand{\ephmean}{\left<\epsilon_\mathrm{ph}\right>}
\newcommand{\gammarad}{\gamma_\mathrm{rad}}

\begin{document}

\title{Synchrotron Pair Production Equilibrium in Relativistic Magnetic Reconnection}

\author[0000-0002-4738-1168]{Alexander Y. Chen}
\affil{Physics Department and McDonnell Center for the Space Sciences, Washington University in St. Louis; MO, 63130, USA}

\author{Dmitri Uzdensky}
\affil{Center for Integrated Plasma Studies, Department of Physics, University of Colorado, 390 UCB, Boulder, CO 80309-0390, USA}

\author{Jason Dexter}
\affil{JILA, University of Colorado and National Institute of Standards and Technology, 440 UCB, Boulder, CO 80309-0440, USA}

\correspondingauthor{Alex Chen}
\email{cuyran@wustl.edu}

\begin{abstract}

    Magnetic reconnection is ubiquitous in astrophysical systems, and in many
    such systems, the plasma suffers from significant cooling due to synchrotron
    radiation. We study relativistic magnetic reconnection in the presence of
    strong synchrotron cooling, where the ambient magnetization $\sigma$ is high
    and the magnetic compactness $\ell_{B}$ of the system is of order unity. In
    this regime, $e^{\pm}$ pair production from synchrotron photons is
    inevitable, and this process can regulate the magnetization $\sigma$
    surrounding the current sheet. We investigate this self-regulation
    analytically and find a self-consistent steady state for a given magnetic
    compactness of the system and initial magnetization. This result helps
    estimate the self-consistent upstream magnetization in systems where plasma
    density is poorly constrained, and can be useful for a variety of
    astrophysical systems. As illustrative examples, we apply it to study the
    properties of reconnecting current sheets near the supermassive black hole
    of M87, as well as the equatorial current sheet outside the light cylinder
    of the Crab pulsar.
\end{abstract}


\section{Introduction}

Magnetic reconnection is a fundamental plasma physics process that dissipates
magnetic energy into plasma heating and nonthermal particle acceleration. Over
the past decade, significant progress in our understanding of collisionless
relativistic reconnection has been made through first-principles
Particle-in-Cell (PIC) simulations
\citep[e.g.][]{2012ApJ...746..148C,2014ApJ...783L..21S,2014PhRvL.113o5005G,
  2016ApJ...816L...8W,2018MNRAS.473.4840W}. This process accelerates particles
into a power-law energy distribution with index $p$ and cutoff
$\gamma_\mathrm{max}$ that depend on the upstream magnetization
$\sigma = B_{0}^{2}/4\pi n_{0}m_{e}c^{2}$ \citep[e.g.][]{2014ApJ...783L..21S}.
In the ultra-relativistic limit of very high $\sigma\gg 1$, the power-law index
$p$ of the accelerated particles approaches unity, $p \to 1$
\citep{2001ApJ...562L..63Z,2008ApJ...682.1436L,2014PhRvL.113o5005G,
  2016ApJ...816L...8W}.

However, the effects of radiation and their feed back to the process of magnetic
reconnection are starting to be studied only recently
\citep[e.g.][]{2009PhRvL.103g5002J,2011PhPl...18d2105U,2011ApJ...737L..40U,2012ApJ...746..148C,2014ApJ...780....3U,2014ApJ...782..104C,2016ASSL..427..473U,2017ApJ...850..141B,2019MNRAS.482L..60W,2019ApJ...870...49S,2019ApJ...877...53H,2020ApJ...899...52S,2020MNRAS.498..799M,2021MNRAS.507.5625S,2021MNRAS.508.4532M}.
In extreme astrophysical environments, particles often suffer from significant
radiative cooling through synchrotron or inverse Compton (IC) radiation. This
can lead to significant changes to the power-law index of the particle energy
spectra \citep{2016ApJ...816L...8W}, and can produce observable intermittency
through kinetic beaming
\citep{2012ApJ...754L..33C,2013ApJ...770..147C,2020MNRAS.498..799M}. In some
systems, the radiated photons can be energetic enough to produce $e^{\pm}$ pairs
in the upstream or close to the reconnecting current sheet, regulating the
plasma supply and the magnetization parameter~$\sigma$. Such systems include
magnetospheres of rotation-powered pulsars (outside the light cylinder)
\citep[e.g.][]{1996A&A...311..172L,2019ApJ...877...53H},
magnetospheres and coronae of accreting black holes
\citep[e.g.][]{2017ApJ...850..141B,2020ApJ...899...52S,2021MNRAS.507.5625S,2021MNRAS.508.4532M}, and active magnetospheres of magnetars \citep{2011SSRv..160...45U,2021ApJ...921...92B}.

For many of these astrophysical systems, the magnetization parameter $\sigma$
can be poorly constrained, due to uncertainties in either the local magnetic
field strength or plasma density. For example, the pair multiplicity injected by
the Crab pulsar into its Pulsar Wind Nebula (PWN) is a long standing problem
\citep[see e.g.][]{2014IJMPS..2860160A}. The plasma density near the light
cylinder can be quite uncertain and is likely regulated by local pair production
\citep{2019ApJ...877...53H,2021arXiv210903935H}. For supermassive black holes,
recent GRMHD simulations have shown that Magnetically Arrested Disks (MAD) can
undergo quasi-periodic eruptions that form transient current sheets in a
low-density region near the black hole horizon
\citep[e.g.][]{2021MNRAS.508.1241C,2022MNRAS.511.3536S,2022ApJ...924L..32R}.
However, due to artificial mass injection in MHD codes, it is difficult to
constrain the magnetization in such an environment from first principles through
simulations. This makes it difficult to make theoretical predictions about the
radiative signatures of these current sheets.

In this paper, we investigate the effect of $e^{\pm}$ pair production from
synchrotron photons in a reconnecting current sheet, and study how the plasma
density self-regulates to an equilibrium. Section~\ref{sec:params} defines the
basic parameters of this problem and our assumptions. Section~\ref{sec:model}
presents an analytic model that captures the basic features of this
self-regulation through pair production. Section~\ref{sec:application} applies
the model to two astrophysical scenarios: the supermassive black hole in M87
(Section~\ref{sec:M87}) and the Crab pulsar (Section~\ref{sec:crab}).
Section~\ref{sec:comparison} compares our results with previous works on
related topics, and finally in Section~\ref{sec:discussion} we discuss some of
the potential shortcomings of this model and possible future extensions.

\section{Basic Parameters}
\label{sec:params}

One of the key dimensionless parameters that govern the magnetic reconnection physics is the upstream plasma magnetization:
\begin{equation}
    \label{eq:sigma}
    \sigma \equiv \frac{B_{0}^{2}}{4\pi \rho_{0}c^{2}},
\end{equation}
where $B_{0}$ is the reconnecting magnetic field strength, and $\rho_{0}c^{2}$
is the rest-mass energy density of the upstream plasma. This definition assumes
that the upstream is cold, or $kT \ll m_{e}c^{2}$, such that the relativistic
enthalpy of the upstream plasma is primarily given by its rest
mass\footnote{Otherwise, one will need to also define the ``hot'' magnetization
  $\sigma_{h}$ using the full enthalpy of the upstream plasma, taking into
  account its relativistic internal energy and pressure.}. This assumption is
appropriate in an environment with strong cooling. We also limit our
consideration to an $e^{\pm}$ plasma, which is appropriate in an environment
where copious pair production is expected. Furthermore, we shall assume that
reconnection proceeds in the relativistic regime, marked by $\sigma\gg 1$,
expected in extreme astrophysical environments around black holes and neutron
stars. Finally, in this paper we will focus on reconnection without guide field,
which is relevant to the applications that we will discuss in
Section~\ref{sec:application}.

Relativistic reconnection-driven nonthermal particle acceleration in this
so-called zero-guide-field case in a pair plasma has been well-studied in many
previous PIC studies, especially in the non-radiative case
\citep[e.g.][]{2004PhPl...11.1151J,2007ApJ...670..702Z,2014ApJ...783L..21S,2014PhRvL.113o5005G,2016ApJ...816L...8W}.
In particular, it was found that particles undergo rapid primary acceleration in
the elementary inter-plasmoid current layers around X-points to a power-law
energy distribution $f(\gamma) \sim \gamma^{-p}$, with a $\sigma$-dependent
power-law index which approaches unity in the ultra-relativistic high $\sigma$
limit, $p(\sigma) \to 1$ as $\sigma \to \infty$ \citep[see
e.g.][]{2001ApJ...562L..63Z,2003ApJ...586...72L,2008ApJ...682.1436L,2018MNRAS.473.4840W}.
This primary power law extends up to around $\gamma_\mathrm{max} \sim 4\sigma$
\citep[see e.g.][]{2016ApJ...816L...8W,2022JPlPh..88a9014U}, perhaps followed by
a steeper higher-energy power-law spectrum
\citep[e.g.][]{2018MNRAS.481.5687P,2021ApJ...912...48H}, and finally by an
exponential cutoff \citep{2008ApJ...682.1436L,2016ApJ...816L...8W}.

Synchrotron cooling introduces a second important dimensionless parameter, the
magnetic compactness~$\ell_{B}$, which measures the radiative energy loss rate
of marginally relativistic ($\gamma \sim 1$) electrons with respect to the light
crossing time of the system:
\begin{equation}
    \label{eq:compactness}
    \ell_{B} \equiv \frac{\sigma_{T}U_{B}L}{m_{e}c^{2}},
\end{equation}
where $\sigma_{T}$ is the Thomson cross section, $U_{B} = B_{0}^{2}/8\pi$ is the
magnetic energy density, and $L$ is the system size, which we take to be the
length of the current sheet.

For a given electron with Lorentz factor $\gamma\gg 1$ gyrating in a magnetic
field with pitch angle $\theta = \pi/2$, its synchrotron cooling time is:
\begin{equation}
    \label{eq:sync_cooling_time}
    t_\mathrm{cool} = \frac{\gamma m_{e}c^{2}}{4\sigma_{T}c\gamma^{2}U_{B}/3}\, .
\end{equation}
Equating this with the light-crossing time of the system, $L/c$, one can find
that the electron will cool to a Lorentz factor of
$\gamma_\mathrm{cool} \sim 1/\ell_{B}$ over the system light-crossing time if
$\ell_{B} \lesssim 1$. If $\ell_{B} > 1$, the electron will become
non-relativistic before it leaves the system. The time for a particle with
arbitrary initial Lorentz factor $\gamma\gg 1$ to cool down to $\gamma \sim 1$
is $t \sim L/c\ell_{B}$. In this paper, we study systems where
$\ell_{B}\gtrsim 1$. In these systems synchrotron cooling is efficient, and
leptons accelerated in the reconnection layer will cool to Lorentz factors
$\gamma \sim 1$ before they exit the system. This also provides reasonable
justification for our assumption that the upstream plasma is relativistically
cold, $kT \ll m_{e}c^{2}$, in our definition of upstream magnetization $\sigma$
[see Equation~\eqref{eq:sigma}].

When dealing with radiative reconnection, it is often beneficial to define a
radiation reaction-limited Lorentz factor by balancing the radiation-reaction
force with the accelerating electric force due to the typical reconnection
electric field $E_\mathrm{rec} = \beta_\mathrm{rec}B_0$:
\begin{equation}
    \label{eq:gamma_rad}
    \gammarad \equiv \sqrt{\frac{3e\beta_\mathrm{rec}B_{0}}{4\sigma_{T}U_{B}}},
\end{equation}
where $\beta_{\rm rec} \simeq 0.1$ is the dimensionless collisionless
relativistic reconnection rate. Since this quantity does not involve the system
size $L$, it measures the \emph{local} relative strength of radiative cooling.
Note that since particles are accelerated near X-points deep inside the current
layer, where the $B$-field is small, their Lorentz factors can exceed
$\gammarad$ locally
\citep{2004PhRvL..92r1101K,2011ApJ...737L..40U,2012ApJ...746..148C}. Only once
the particles encounter regions of strong perpendicular magnetic field, e.g.\
when they are captured in plasmoids, will they start radiating away most of
their energy. The critical energy $\gammarad$ is related to $\ell_{B}$ by the
system size:
\begin{equation}
    \label{eq:gamma_rad_lB}
    \frac{4}{3}\gammarad^{2}\ell_{B} = 
    \beta_\mathrm{rec}\frac{L}{\rho_{0}},
\end{equation}
where $\rho_{0} = m_{e}c^{2}/eB_{0}$ is the nominal relativistic electron Larmor radius in the upstream field.

Finally, given a characteristic magnetic field $B_{0}$ and electron Lorentz factor~$\gamma$, the characteristic synchrotron photon frequency is:
\begin{equation}
    \label{eq:sync_energy}
    \omega_{c} = \frac{3}{2}\gamma^{2}\omega_{B}\sin\theta,
\end{equation}
where $\omega_{B} = eB_{0}/m_{e}c$ is the standard non-relativistic electron
cyclotron frequency and $\theta$ is the particle's pitch angle. The synchrotron
photon will be capable of creating an $e^{\pm}$ pair when its energy is larger
than~$m_{e}c^{2}$, or in other words, when the Lorentz factor of the emitting
particle is larger than
\begin{equation}
    \label{eq:gamma_c}
    \gamma_{c} = \sqrt{\frac{2B_{Q}}{3B_{0}\sin\theta}},
\end{equation}
where $B_{Q} \equiv m_{e}^{2}c^{3}/e\hbar = 4.4\times 10^{13}\,\mathrm{G}$ is
the quantum critical (Schwinger) magnetic field. If the plasma magnetization
$\sigma \gtrsim \gamma_{c}/4$, so that $\gamma_{\rm max} \gtrsim \gamma_c$, then
there will be an appreciable number of reconnection-accelerated particles
capable of radiating photons that can convert to pairs through photon-photon
collisions.

\section{Analytic Model}
\label{sec:model}

Consider a reconnecting current sheet with initial
$\sigma_{0} > \gamma_{c} \gg 1$ and thus capable of pair production through
synchrotron photons. We are interested in the regime where synchrotron cooling
is efficient, $\ell_{B}\gtrsim 1$. We expect that once reconnection begins,
synchrotron photons start to create pairs in the vicinity of the current sheet,
producing extra plasma that will regulate the effective upstream magnetization
$\sigma$, reducing it from its initial, far-upstream value~$\sigma_{0}$. The
effective magnetization may then become smaller than~$\gamma_{c}$, and this
would strongly suppress pair production since particles would no longer be able
to gain sufficient energy to emit pair-producing photons. This may lead to a
limit cycle behavior, as was noted by \citet{2021MNRAS.508.4532M} in the case of
inverse-Compton radiation; this behavior is somewhat similar to a pair-producing
gap in the magnetospheres of black holes and neutron stars
\citep[e.g.][]{2020ApJ...895..121C,2020ApJ...902...80K}. Alternatively, the
system may be able to self-regulate to a quasi-steady state with an
equilibrium~$\sigma_{\pm}$. In this section, we quantitatively describe this
process using a simple analytic model, and evaluate whether an asymptotic state
will be reached.

Given an (assumed to be isotropic) photon distribution
$n_{\gamma}(\epsilon) = n_{\gamma}f_{\gamma}(\epsilon)$, where $\epsilon = \hbar\omega_\mathrm{ph}/m_ec^2$ is the dimensionless photon energy and $f_\gamma$ is normalized to unity, the pair production rate can be calculated as:
\begin{equation}
    \label{eq:pair_rate}
    \dot{n}_{\pm} = \int n_{\gamma}f_{\gamma}(\epsilon) \nu_{\gamma\gamma}(\epsilon)\,d\epsilon,
\end{equation}
where $\nu_{\gamma\gamma}$ is the $\gamma$--$\gamma$ pair production rate for a
single photon of energy $\epsilon$ \citep{1967PhRv..155.1404G}:
\begin{equation}
    \label{eq:gg_rate}
    \nu_{\gamma\gamma}(\epsilon) = \iint \frac{1}{2}n_{\gamma}(\epsilon')\sigma_{\gamma\gamma}c(1 - \cos\theta)\sin\theta\,d\theta d\epsilon',
\end{equation}
where $\sigma_{\gamma\gamma}$ is the total collision cross section for two
photons of energies $\epsilon$ and $\epsilon'$ with relative angle~$\theta$. We
assume magnetic reconnection produces a power-law photon spectrum:
\begin{equation}
    f_{\gamma} = A \epsilon^{-(\alpha+1)}e^{-\epsilon/\emax},
    \label{eq:ph_spectrum}
\end{equation}
which extends from $\emin$ to $\emax$ with an exponential cutoff above $\emax$.
The pair production rate can then be written as
\citep{1987MNRAS.227..403S}\footnote{This expression was derived without the
  exponential cutoff. However, in our application we mainly use the pair
  production rate when $\epsilon > 1$, and $n(1/\epsilon)$ samples the low
  energy part of the spectrum, insensitive to the exponential cutoff. Therefore
  we use this expression directly as an approximation.}:
\begin{equation}
    \label{eq:power_law_gg_rate}
    \nu_{\gamma\gamma}(\epsilon) \simeq c\sigma_{T}\eta(\alpha)\frac{n(1/\epsilon)}{\epsilon},
\end{equation}
where $\eta(\alpha)$ has an approximate form:
\begin{equation}
    \eta(\alpha) \approx \frac{7}{6}(2+\alpha)^{-1}(1+\alpha)^{-5/3}.
\end{equation}
The pair production rate is dominated by the interaction of photons of energies
$\epsilon$ and~$1/\epsilon$. By integrating over the whole photon distribution,
equation~\eqref{eq:pair_rate} double-counts all potential pair-producing energy
combinations, therefore it is useful to set the lower limit of the integration
to~1:
\begin{equation}
\begin{split}
    \dot{n}_\pm &\simeq n_\gamma^2c\sigma_T\eta(\alpha)A^2\int_1^{\infty}\frac{1}{\epsilon}e^{-\epsilon/\emax}\,d\epsilon \\
    &= n_\gamma^2c\sigma_T\eta(\alpha)A^2 E_{1}(1/\emax),
\end{split}
\label{eq:pair_rate_integral}
\end{equation}
where $E_{1}$ is the exponential integral defined as
$E_{1}(z) = \int_{z}^{\infty}t^{-1}e^{-t}\,dt$. It scales as $\log(\emax)$ when
$\emax \gg 1$, and drops to zero exponentially when $\emax \ll 1$. Physically
this means that $\dot{n}_{\pm}$ is exponentially suppressed when only the
photons with energies significantly above the exponential cutoff are capable of
producing pairs. Introducing an exponential cutoff allows us to gracefully
handle the transition from $\emax > 1$ to $\emax < 1$. If we were to adopt a
power-law distribution with a sharp cutoff at~$\emax$, then $\dot{n}_{\pm}$
would go to zero as soon as $\emax$ drops below unity, which may have led to
unrealistic conclusions. We have also assumed that
$1/\emin \gg \emax \gg \emin$, such that the upper limit in the integral in
Equation~\eqref{eq:pair_rate_integral} can be effectively taken to be infinity
instead of $1/\emin$. This assumption holds for the astrophysical applications
that we explore in this paper.

We can estimate $n_{\gamma}$ in general terms as follows. Consider reconnection as a
process that converts a portion of magnetic energy into particle energy and
eventually into radiation. The energy flux of photons away from the reconnection
layer should equal to a certain fraction of the dissipated magnetic energy:
\begin{equation}
    \label{eq:n_gamma}
    \ephmean m_ec^2n_{\gamma}c\sin\theta_b = \kappa S = 2\kappa \beta_\mathrm{rec}c\frac{B_{0}^{2}}{4\pi} = 4\kappa \beta_\mathrm{rec}cU_{B},
\end{equation}
where $S$ is the Poynting flux into the current sheet, $\ephmean$ is the mean
photon energy normalized to~$m_e c^2$, and $\kappa$ is an efficiency factor that
quantifies how much magnetic energy is converted to synchrotron radiation. In a
radiatively efficient system, most of the dissipated magnetic energy is radiated
away, hence $\kappa$ essentially measures the amount of energy that is given to
the plasma. PIC simulations have suggested that $\kappa \sim 0.5$ \citep[see
e.g.][]{2020ApJ...899...52S}, and we will adopt this fiducial value for our
estimates. Next, the factor $\sin\theta_b$ accounts for the fact that
synchrotron emission may have a degree of beaming with respect to the current
sheet. We adopt $\sin\theta_{b}\simeq 0.5$ in our model, but acknowledge that it
may be lower in reality. The leading factor of 2 accounts for incoming Poynting
flux from both above and below the current sheet. Finally, $\beta_\mathrm{rec}$
is the normalized reconnection speed, which has been measured to be close to
$\beta_\mathrm{rec}\sim 0.1$ in PIC simulations of relativistic reconnection.

Given the photon distribution described by Equation~\eqref{eq:ph_spectrum}, we
can directly compute~$\ephmean$:
\begin{equation}
    \begin{split}
    \ephmean &= A\int_{\emin}^{\infty}\epsilon^{-\alpha}e^{-\epsilon/\emax}\,d\epsilon \\
             &\simeq A\left[\emax^{-\alpha+1}\Gamma(1-\alpha)-\frac{\emin^{-\alpha+1}}{1-\alpha} + O\left(\frac{\emin}{\emax}\right)^{2}\right].
    \end{split}
\end{equation}
For $\alpha < 1$ and $\emax \gg \emin$, the expression simplifies to
$\ephmean \approx A\emax^{1-\alpha} \, \Gamma(1 - \alpha)$. In other words, the
photon spectrum is dominated by the high-energy end. In the opposite regime
where $\alpha > 1$, the spectrum is instead dominated by the low-energy end, and
$\ephmean \approx A \emin^{\alpha - 1}/(\alpha - 1)$. In realistic astrophysical
systems where cooling is efficient, $\emin$ can either approach the photon
energy corresponding to the electron cyclotron frequency, or be regulated by
synchrotron self-absorption. On the other hand, $\emax$ is determined by the
nonthermal particle acceleration mechanism. The normalization constant $A$ will
eventually cancel out with the same factor in
Equation~\eqref{eq:pair_rate_integral}.

Recent progress on PIC simulations of relativistic reconnection can inform us
about the dependence of $\emax$ and $\alpha$ on the upstream magnetization
$\sigma$, since this synchrotron photon field is produced by the nonthermal
particles accelerated in the current sheet. As discussed in
Section~\ref{sec:params}, it has been shown that in the ultra-relativistic limit
of very high $\sigma \gg 1$, the particles are promptly accelerated to a hard
power-law distribution of index $p \sim 1$. This is the limit that is
appropriate for the magnetospheres of compact objects, which we are ultimately
interested in (see Section~\ref{sec:application}). However, since all leptons in
the system will be fast-cooling if $\ell_B \gtrsim 1$, the cooled particle
spectrum becomes $p \sim 2$ and the radiation spectrum then has a power-law
index $\alpha = (p - 1)/2 \sim 0.5$. In reality, the instantaneous particle
spectrum in radiative reconnection is likely highly variable \citep[see
e.g.][]{2019MNRAS.482L..60W,2019ApJ...877...53H}, but the overall radiation
spectrum is dominated by the times when the particle spectrum is hardest. Thus,
$\alpha \sim 0.5$ can be a good approximation to the time-averaged photon
spectrum, and we shall adopt this value in our analysis. In this limit, the mean
photon energy becomes:
\begin{equation}
    \label{eq:ephmean-limit}
    \ephmean \simeq 1.77A\emax^{0.5} \, .
\end{equation}

The maximum extent of the power-law photon distribution $\emax$ is directly
determined by the maximum extent $\gamma_\mathrm{max}$ of the particle energy
power law,
\begin{equation}
    \emax = \gamma_\mathrm{max}^2b\sin\theta,
\end{equation}
where $b \equiv B_{0}/B_{Q}$ is the dimensionless ratio of the upstream magnetic
field to the Schwinger magnetic field, and $\theta$ is the average pitch angle
of the particle distribution. For simplicity we take a typical value of
$\sin\theta = 1/2$. How the power-law cutoff $\gamma_\mathrm{max}$ depends on
the reconnection physics is still an actively debated issue. The first serious
study of this maximum extent of the power-law distribution was conducted by
\citet{2016ApJ...816L...8W}, who found that $\gamma_\mathrm{max} \sim 4\sigma$,
above which the particle distribution transitions to an exponential cutoff. More
recently, \citet{2018MNRAS.481.5687P} and \citet{2021ApJ...912...48H} found that
over time the system may develop a secondary power law above $4\sigma$ due to
plasmoid compression; however, this effect was only demonstrated in 2D and it is
not clear whether it persists in strongly radiative environments. In this paper
we will adopt $\gamma_\mathrm{max} = 4\sigma$ and come back to this issue in
Section~\ref{sec:discussion}. Under this assumption, the maximum extent of the
power-law photon distribution can be written as $\emax \simeq 8\sigma^{2}b$.

We can now introduce $\ephmean$ and $\emax$ back into
Equation~\eqref{eq:pair_rate} to recover an equation that only depends on the
physical parameters of the reconnection layer and the upstream
magnetization~$\sigma$:
\begin{equation}
    \begin{split}
      \dot{n}_{\pm} &= 16\kappa^{2}\beta_\mathrm{rec}^{2}\frac{U_{B}}{m_{e}c^{2}}\ell_{B}\frac{c}{L}\frac{\eta(\alpha)A^2E_{1}(1/\emax)}{\ephmean^{2}\sin^2\theta_b}, \\
                    &\simeq 1.6\times 10^{-3}\frac{U_{B}}{m_{e}c^{2}}\ell_{B}\frac{c}{L}\frac{E_{1}\left[(8\sigma^{2}b)^{-1}\right]}{\sigma^{2} b},
    \end{split}
\end{equation}
where we have made the substitutions $\kappa \approx 0.5$,
$\beta_\mathrm{rec} \approx 0.1$, $\alpha\approx 0.5$, and
$\sin\theta_{b}\approx 0.5$.

The balance between pair production and escape will determine the equilibrium plasma density and upstream magnetization. The pairs produced
  through collisions of synchrotron photons will in general escape in two ways:
  they either stream along the upstream magnetic field at the speed of light and
  exit the system, or they drift into the reconnecting current sheet at
  $v_\mathrm{rec} \simeq 0.1c$ and participate in the reconnection process.
If pairs are predominately produced at a distance $d$ from the current
    sheet that is larger than $\beta_\mathrm{rec}L$, then they will tend to
    escape the system before drifting into the reconnection layer, whereas if
    $d < \beta_\mathrm{rec}L$ the pairs will escape through drifting into the
    current sheet. In an environment with magnetic compactness $\ell_{B}\sim 1$,
    the characteristic optical depth to pair production can be estimated as:
\begin{equation}
    \label{eq:pair_optical_depth}
    \tau_\mathrm{ph} = \frac{L}{l_\mathrm{ph}} \sim n_{\gamma}\sigma_{T}L \sim \beta_\mathrm{rec}\ell_{B} < 1.
\end{equation}
As a result, $e^{\pm}$ pairs are typically produced far away from the current
sheet and escape the system at the speed of light. This is the assumption that
we will adopt in this analytic model. Under this assumption, the equilibrium
number density of $e^{\pm}$ pairs is simply $n_{\pm}\simeq \dot{n}_{\pm}L/c$.
This equilibrium density defines an effective magnetization via the
equation\footnote{Here we have also assumed that the density of created pairs is
  much higher than the ambient pair density (which thus becomes irrelevant),
  dominating the final plasma density. This assumption holds in the applications
  examined in Section~\ref{sec:application}.}:
\begin{equation}
    \begin{split}
      \sigma_{\pm} &\sim \frac{U_{B}}{n_{\pm}m_{e}c^{2}} = \frac{\sin^2\theta_b}{16\kappa^{2}\beta_\mathrm{rec}^{2}\ell_{B}\eta(\alpha)A^2}\frac{\ephmean^{2}}{E_{1}(1/\emax)} \\
                   &\simeq 6\times 10^{2}\, \frac{\sigma_{\pm}^{2}b}{\ell_{B} E_{1}\left[(8\sigma_{\pm}^{2}b)^{-1}\right]}.
    \end{split}
    \label{eq:sigma_eff}
\end{equation}
Equation~\eqref{eq:sigma_eff} is a transcendental algebraic equation for the
effective magnetization $\sigma_{\pm}$ after synchrotron pair production has
come to an equilibrium. In essence, we are looking for a pair-production
equilibrium in a system where the number of pairs created directly correlates
with the efficiency of nonthermal acceleration. Magnetic reconnection is an
example of such a system, where the maximum particle acceleration correlates
with~$\sigma$, which in turn is determined by the numbers of $e^{\pm}$ pairs
produced.

\begin{figure}[t]
    \centering
    \includegraphics[width=0.47\textwidth]{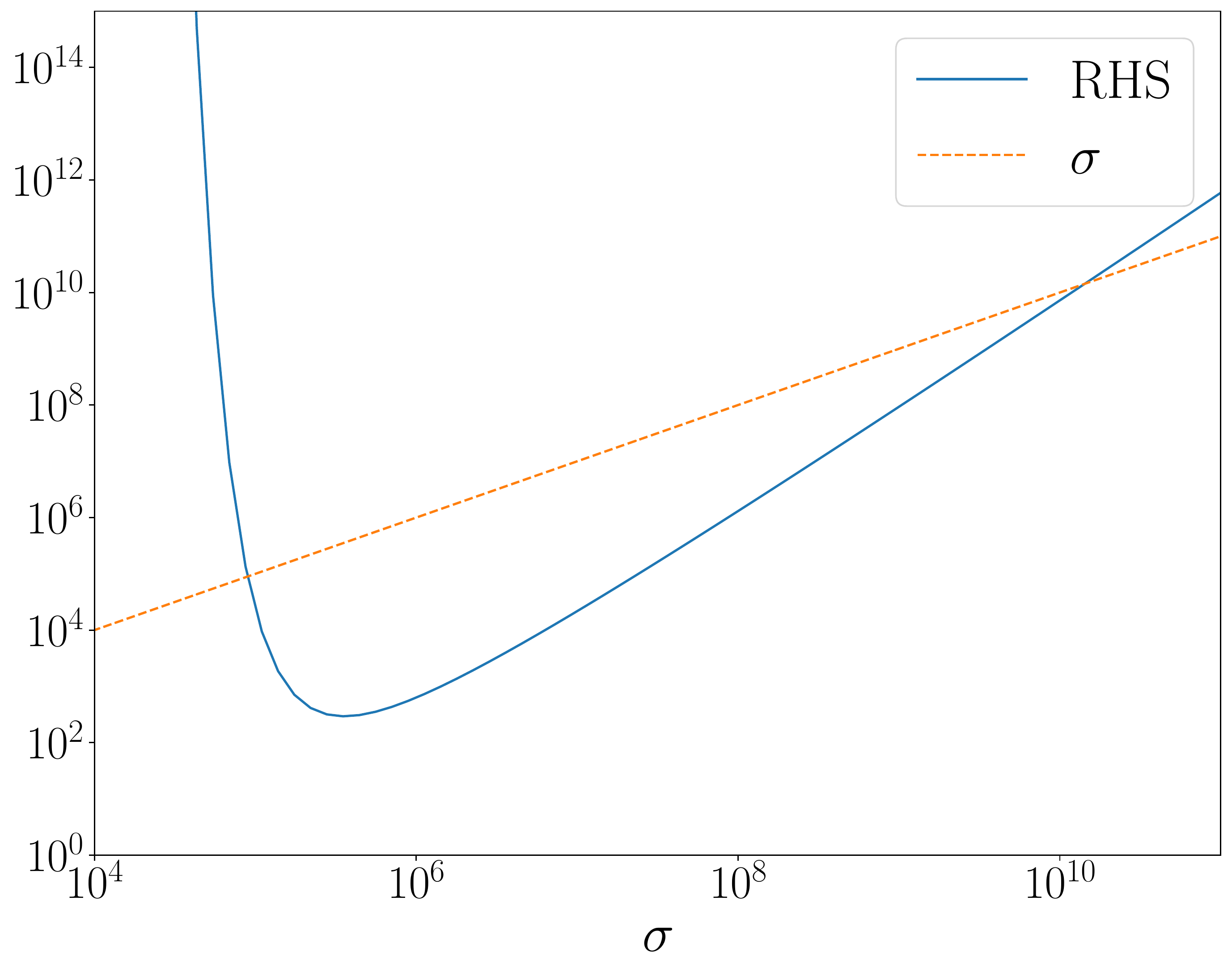}
    \caption{Example left and right hand sides of Equation~\ref{eq:sigma_eff},
      evaluated for $\ell_{B} = 1$, $\alpha = 0.5$, and $B = 100\,\mathrm{G}$.
      There are two solutions for $\sigma$ where the two curves cross. The lower
      solution is stable while the upper solution is unstable.}
    \label{fig:lhs-v-rhs}
\end{figure}

Figure~\ref{fig:lhs-v-rhs} illustrates the general behavior of
Equation~\eqref{eq:sigma_eff}. The right-hand side is a convex function and goes
to infinity both when $\sigma\to 0$ and $\sigma\to \infty$. There are in general
two solutions of the equation when the RHS crosses identity. The lower solution
occurs close to the exponential suppression of the pair production rate, where
$\gamma_\mathrm{max}\sim \gamma_{c}$, while the upper solution occurs at a much
larger~$\sigma$, such that $\emax \gg 1$. In the latter limit,
$E_{1}(1/\emax)\to \log(\emax)$, and the solution of
Equation~\eqref{eq:sigma_eff} approaches $\sigma_{\pm}\sim \ell_{B}/600 b$.

However, the upper solution may not always be physical. Taking the result in
Figure~\ref{fig:lhs-v-rhs} for example, the upper solution is close to
$\sigma\sim 1.4\times 10^{10}$, at which point $\emax \sim 3.6\times 10^{9}$,
far higher than the synchrotron burnoff limit of 160~MeV. In general, this
solution is self-consistent only when the resulting $\sigma$ is not too far
above~$\gammarad$, so that the high-energy extent of the particle power law is
controlled by $\sigma$ and not by radiative cooling. Even when this upper
solution is within the physical applicability of the model, it is still
unstable: increasing $\sigma$ from there decreases pair production rate
[Equation~\eqref{eq:pair_rate}], triggering further increase of $\sigma$ until
reaching the background~$\sigma_{0}$. This is because above this magnetization
the hard radiation spectrum implies that there are not enough low energy targets
for the high-energy synchrotron photons to pair-create on. Decreasing $\sigma$
from the upper solution, on the other hand, increases pair production rate and
further decreases $\sigma$ from the freshly generated $e^{\pm}$ plasma.

On the other hand, the lower solution is stable, and it is the solution that we
seek. As mentioned above, this solution typically arises when
$\gamma_\mathrm{max}$ is close to the pair-production threshold Lorentz factor
$\gamma_{c}$ [c.f. Equation~\eqref{eq:gamma_c}], and further decreasing $\sigma$
causes an exponential suppression in the pair production rate. As long as the
system starts off with a magnetization $\sigma_{0}$ between the two solutions,
it will be driven towards the stable lower solution by self-regulated
synchrotron pair production. Even though the initial $\sigma_{0}$ may be higher
than the radiation-limited Lorentz factor~$\gammarad$, we find that this final
equilibrium $\sigma_{\pm}$ is much smaller than $\gammarad$ for a wide range of
magnetic field strengths.

\begin{figure}[h]
    \centering
    \includegraphics[width=0.48\textwidth]{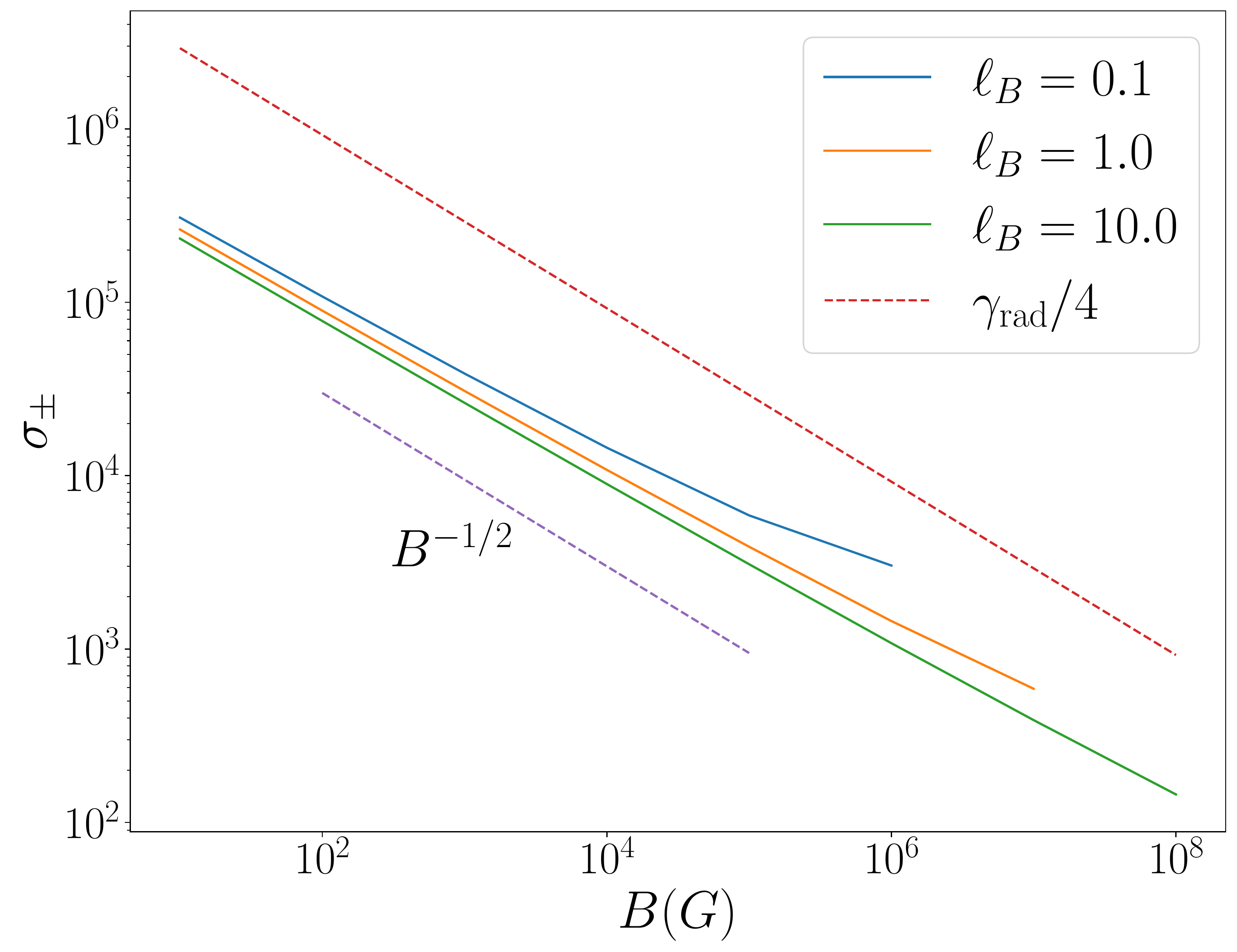}
    \caption{Dependence of the lower solution $\sigma_\pm$ on the upstream
      magnetic field $B$ at different magnetic compactness~$\ell_B$, at
      $\alpha = 0.5$. For a wide range of magnetic field strengths, this
        solution is below the radiation limited Lorentz factor~$\gammarad$.
      Note that with increasing $B$ field, the two solutions become closer and
      beyond certain point there is no longer a solution to
      Equation~\eqref{eq:sigma_eff}, which is close to where the curves terminate. For smaller compactness~$\ell_{B}$, this transition occurs at a lower magnetic field strength.}
    \label{fig:sigma-vs-B}
\end{figure}

Figure~\ref{fig:sigma-vs-B} shows how the self-regulated magnetization
$\sigma_\pm$ (lower solution) depends on upstream magnetic field $B$ and the
system compactness. The solution is obtained numerically using Newton's method.
For a given magnetic compactness, Equation~\eqref{eq:sigma_eff} ceases to have a
solution when $B$ is large enough, as its RHS no longer intersects with the
identity line on Figure~\ref{fig:lhs-v-rhs}. In general, the two solutions
approach each other at higher $B$-fields. One can also see from
Figure~\ref{fig:sigma-vs-B} that the equilibrium $\sigma_\pm$ depends much more
sensitively on the magnetic field strength $B$ than on the system size which
manifests as~$\ell_B$. Qualitatively, this is because the lower solution is
always pushed close to the pair production threshold $\gamma_{c}$ due to
exponential suppression of pair production rate below it, and $\gamma_{c}$
scales as~$b^{-1/2}$ [c.f. Equation~\eqref{eq:gamma_c}]. In contrast, the
equilibrium magnetization depends weakly
on other parameters such as $\alpha$ and~$\ell_{B}$.

Note however that these results rely on the assumption that $\alpha < 1$, so
that the photon spectrum peaks at the high-energy end. At high $\ell_{B}\gg 1$
or very strong cooling, this assumption may break; unfortunately, both the
reconnection-accelerated particle spectrum and the radiation spectrum in this
regime have not yet been sufficiently well studied in first-principles radiative
PIC simulations and are thus still poorly understood. On the other hand, when
$\ell_{B}\ll 1$, there is a spectral break for the synchrotron spectrum at low
energies since particles with Lorentz factors $\gamma < 1/\ell_{B}$ do not cool
appreciably before leaving the system. The spectral break changes the target
photon distribution, and our simple assumption of
Equation~\eqref{eq:power_law_gg_rate} needs to be replaced. Our analytic model
proposed in this section works best when $\ell_{B}$ is not too far from unity.

\section{Astrophysical Applications}
\label{sec:application}

We will now apply the analytic model described in Section~\ref{sec:model} to two
specific astrophysical scenarios where this process may prove to be relevant: a
reconnecting current sheet in the magnetosphere of the central supermassive
black hole of~M87, and the equatorial current sheet outside the light cylinder
of the Crab pulsar.

\subsection{The Magnetosphere of M87}
\label{sec:M87}

Recent GRMHD models and their comparison with the observations of the Event
Horizon Telescope (EHT) have significantly improved our understanding of the
structures of accretion flow and the magnetic field close to the event horizon
of the supermassive black hole at the center of M87 \citep{2019ApJ...875L...1E}.
Magnetically Arrested Disk (MAD) models seem to be favored by the recent
polarization measurements \citep{2021ApJ...910L..13E}. These MAD models tend to
predict a magnetic field of $B\sim 1\text{--}30\,\mathrm{G}$ in the
millimeter-emission region. Depending on the numerical models, the magnetic
field at the horizon can be as high as $B\sim 20\text{--}200\,\mathrm{G}$
\citep{2021MNRAS.507.4864Y,2022ApJ...924L..32R}. At the same time, GRMHD
simulations of the MAD model tend to observe large-scale current sheets that can
form in the equatorial plane near the event horizon during an eruption
\citep[see,
e.g.][]{2020MNRAS.497.4999D,2021MNRAS.502.2023P,2022MNRAS.511.3536S,2021MNRAS.508.1241C,2022ApJ...924L..32R}.
High-resolution simulations show that these current sheets can be
tearing-unstable and undergo magnetic reconnection, breaking up into a
self-similar plasmoid chain \citep{2022ApJ...924L..32R}. The length scales of
these current sheets can be comparable to the black hole gravitational
radius~$r_g$. Taking $B\sim 200\,\mathrm{G}$ and
$L \sim r_g \sim 10^{15}\,\mathrm{cm}$, we can estimate the system's magnetic
compactness to be $\ell_B \sim 1.3$; therefore synchrotron pair production can
be an important factor that regulates local magnetization. The threshold Lorentz
factor for synchrotron pair production under these conditions is
$\gamma_c \sim 4\times 10^5$ [Equation~\eqref{eq:gamma_c}], and the
radiation-reaction-limited Lorentz factor is
$\gamma_\mathrm{rad} \sim 2\times 10^6$ [Equation~\eqref{eq:gamma_rad}].

For these current sheet parameters, the two roots of
Equation~\eqref{eq:sigma_eff} are $\sigma_{1} \approx 6\times 10^{4}$ and
$\sigma_{2} \approx 2\times 10^{10}$. The initial magnetization $\sigma_0$ in
the magnetic bubble formed in an eruption of the MAD disk is very poorly
constrained. Since the bubble material is torn from the low-density jet funnel,
an upper limit for $\sigma_{0}$ can be estimated using the minimum
Goldreich-Julian charge density which screens local electric field
\citep{1969ApJ...157..869G}. For M87 near the horizon,
$\sigma_\mathrm{GJ} \sim 10^{13}$ \citep{2021MNRAS.507.4864Y}. However, it is
unrealistic to expect that the plasma density is simply characterized by the
Goldreich-Julian charge density, since Comptonized photons from the accretion
disk can produce $e^{\pm}$ pairs in the jet funnel as well as in the vicinity of
the horizon. This process is often called the ``pair drizzle''. \citet{wong2020}
recently calculated the pair production rate from this drizzle mechanism and
estimated that for M87 parameters
$\dot{n}_{\pm} \sim 10^{4}n_\mathrm{GJ}\,c/r_{g}$, which implies a much lower
initial magnetization $\sigma_{0} \sim 10^{9}$. Since this falls below our upper
solution~$\sigma_{2}$, our model predicts that pair production from synchrotron
photons emitted by the particles accelerated in the reconnecting current layer
will lower the magnetization to an equilibrium value of
$\sigma_{\pm} \sim 6\times 10^{4}$, which puts
$\gamma_\mathrm{max}\sim 4\sigma_\pm$ close to $\gamma_c$ and much lower
than~$\gamma_\mathrm{rad}$. This also implies a pair multiplicity of
$\mathcal{M}=n_\pm/n_\mathrm{GJ}\sim 2\times 10^{8}$ over the minimum
Goldreich-Julian density. This result is similar in magnitude to what was
estimated by \citet{2022ApJ...924L..32R} and \citet{2022arXiv220801882K}.
However, our model predicts that the upstream magnetization will stabilize
around this equilibrium value~$\sigma_{\pm}$, which has implications for the M87
VHE gamma-ray flares.

It was proposed by \citet{2022ApJ...924L..32R} that the MAD eruption events that
form large-scale reconnecting current sheets may be a promising mechanism for
powering the TeV flares observed from~M87. However, our results disfavor this
proposition, since synchrotron pair production will quickly lower the
magnetization from its uncertain initial value to $\sim 6\times 10^{4}$, which
limits the extent of the power-law energy distribution to
$\gamma_\mathrm{max}\sim 3\times 10^{5}$. Electrons at this Lorentz factor are
energetically incapable of producing TeV gamma-rays, even if Compton scattering
occurs in the deep Klein-Nishina regime. As a result, it is difficult for the
system to produce a single power-law radiation spectrum that extends from
several~GeV up to several~TeV. Since the equilibrium $\sigma_\pm$ scales
approximately as $B^{-1/2}$ (Figure~\ref{fig:sigma-vs-B}), a significantly lower
magnetic field of $B\sim 2\,\mathrm{G}$ near the horizon is required to increase
$\sigma_{\pm}$ to $6\times 10^{5}$, which would then allow the particle energy
power law to extend beyond a few~TeV. However, such a low magnetic field would
lead to a much lower dissipation power through the reconnection process:
$L_\mathrm{rec}\sim U_Br_g^2\beta_\mathrm{rec}c\sim 5\times 10^{38}\,\mathrm{erg/s}$.
This is much lower than the isotropic equivalent luminosity observed in M87 TeV
flares, which can reach up to $\sim 10^{42}\,\mathrm{erg/s}$
\citep{2012ApJ...746..151A}.

Our synchrotron model does not take into account pairs produced by gamma-ray
photons from inverse-Compton scattering by the accelerated particles. These
gamma-ray photons can be emitted through either synchrotron self-Compton (SSC)
mechanism or by electrons scattering the ambient lower energy photons from the
accretion disk. However, additional channels of pair production will only lower
the final equilibrium magnetization, since there are more ways to generate
plasma. This will, in general, make it even more difficult for the current sheet
to produce TeV emission after pair production kicks in. We will discuss more
about the potential role of SSC photons in Section~\ref{sec:discussion}.

\subsection{Crab Pulsar}
\label{sec:crab}

At the light cylinder of the Crab pulsar, the magnetic field is approximately
$B_\mathrm{LC} \sim 4\times 10^{6}\,\mathrm{G}$ \citep{2014ApJ...780....3U}.
This is computed from the spindown-inferred dipole moment and the rotation
period of the pulsar. The size of the current sheet is given by the
characteristic length scale which is the light cylinder radius
$R_\mathrm{LC} \sim 1.6\times 10^{8}\,\mathrm{cm}$. The resulting system
compactness is $\ell_{B} \sim 30$, close to the regime discussed in
Section~\ref{sec:model}. The nominal magnetization $\sigma_{0}$ at the light
cylinder is mostly determined by the copious $e^\pm$ outflow from the polar cap,
and is generally believed to be $10^3$--$10^5$ \citep{2019ApJ...877...53H}.
Fortunately, our analytic model is insensitive to this initial magnetization, as
long as it lies within the two roots of Equation~\eqref{eq:sigma_eff}. For the
parameters quoted above, the two roots are $\sigma_{1} \approx 5\times 10^{2}$
and $\sigma_{2} \approx 9\times 10^{6}$. This implies that if the current sheet
starts with a magnetization within this range, synchrotron pair production will
drive $\sigma$ towards the lower root which is $\sigma_{\pm} \approx 500$.

The Crab pulsar was observed to emit pulsed VHE gamma-rays that form a power law
extending up to TeV energies \citep{2016A&A...585A.133A}. This puts the
properties of its pulsed emission close to that of~M87. Our predicted low
$\sigma_{\pm}\sim 500$ would seemingly rule out the possibility that pairs
accelerated in the equatorial current sheet can emit TeV gamma-rays, even with a
relativistic bulk-flow boost of $\Gamma\sim 100$. However, only a small fraction
of the spindown power of the Crab is needed to power the pulsed VHE emisssion,
whereas in M87 the luminosity of VHE gamma-ray flares can be comparable to the
jet power. It may be possible that these gamma-rays are produced before the
equilibrium $\sigma_\pm$ is reached.

Outside the light cylinder of the Crab pulsar, the current sheet feeds off the
Y-point and always starts with plasma flowing from the inner magnetosphere with
initial pair density $n_0 \gg n_\mathrm{GJ}$. This flow determines the initial
magnetization $\sigma_0$ surrounding the current sheet that is much higher than
our predicted~$\sigma_\pm$. As pair production kicks in, the outflow
magnetization will gradually drop and stabilize at around the self-regulated
equilibrium $\sigma_{\pm}$ at some distance downstream of the Y-point. Since the
gamma-ray spectrum from GeV to TeV appears to be a single power law
\citep{2016A&A...585A.133A}, we argue that almost all of these gamma-rays are
produced during this time, before the pair equilibrium is established. One can
estimate a characteristic time scale for reaching this pair equilibrium by
computing the time it takes for the initial plasma density to double, normalized
to the system light crossing time:
\begin{equation}
    \label{eq:equilibrium-time}
    \tau_{\pm} = \frac{n_{0}/\dot{n}_{\pm}}{L/c}.
\end{equation}
One can evaluate this time scale using the pair production rate
\eqref{eq:pair_rate} and the initial magnetization~$\sigma_{0}$. For Crab
parameters, if we take an initial $\sigma_{0}\sim 10^{6}$ which is required for
particle acceleration up to~TeV, this characteristic time scale is
$\tau_{\pm} \sim 0.15$. One can take this dimensionless scale as the
characteristic fraction of energy dissipated in the current sheet that can be
emitted in HE to VHE gamma-rays, since pair production will quickly lower
$\sigma$ such that the electron power law only extends to about a~GeV. After
pair production equilibrium is established, the synchrotron spectrum from these
pairs only extends up to about~MeV, which is a far cry from the observed pulsed
gamma-ray component extending from $100\,\mathrm{MeV}$ to $1.5\,\mathrm{TeV}$.
If the current sheet dissipates about 10\% of the spindown power $L_\mathrm{sd}$
between $R_\mathrm{LC}$ and~$2R_\mathrm{LC}$ \citep{2020A&A...642A.204C}, then
the expected gamma-ray luminosity
$L_{\gamma}\lesssim 0.1\tau_{\pm}L_\mathrm{sd}$. Further accounting for the
radiation spectrum and efficiency, this is close to the observed gamma-ray
efficiency of the Crab pulsar of $L_{\gamma}/L_\mathrm{sd}\sim 10^{-3}$
\citep{2013ApJS..208...17A}. A more detailed study on the pulsar gamma-ray
spectrum, efficiency and how it scales with the model parameters is beyond the
scope of this paper, and will be deferred to a future work.

\section{Comparison with Previous Work}
\label{sec:comparison}

Relatively few works have considered the effect of synchrotron pair production
in the process of magnetic reconnection. \citet{1996A&A...311..172L} pointed out
that the reconnecting current sheet outside the light cylinder can potentially
power the high energy emission from gamma-ray pulsars. He also calculated the
number density of the $e^\pm$ plasma that would result from synchrotron pair
production, and concluded that pair production will lower the magnetization
$\sigma$ surrounding the current sheet to the point where no gamma-rays will be
emitted. However, the detailed reconnection-driven particle acceleration
mechanisms were not clearly understood at that time, and the present paper takes
advantage of the recent development of our understanding of the magnetic
reconnection process.

\citet{2019ApJ...877...53H} performed PIC simulations with self-consistent
synchrotron cooling and photon-photon pair production in the context of the Crab
pulsar magnetosphere. However, their simulations were of a very limited
parameter range and at low $\ell_{B}\sim 10^{-2}$. They derived a crude analytic
model to predict the final number of pairs produced from the reconnecting
current sheet, but did not consider the feedback of pair loading on the particle
acceleration process itself. For Crab parameters, they found that the pair
multiplicity $\eta$, defined as the ratio between the final number density and
the initial upstream number density, can become as large as $10^{5}$--$10^{6}$,
which is greater than the initial magnetization $\sigma_0$ at the light
cylinder, indicating that the final $\sigma$ will become less than unity after
pair production saturates. This implies that pair loading should strongly affect
the particle acceleration process and in turn limit the final multiplicity.
This is precisely the problem that we address in the present paper.

\citet{2017ApJ...850..141B,2021ApJ...921...92B} considered magnetic reconnection
in the strongly radiative regime, with synchrotron pair production and inverse
Compton emission in the context of the coronae of X-ray binaries, magnetar
bursts, and electromagnetic precursors of binary neutron star mergers. His works
focused more on the high magnetic compactness regime $\ell_{B} \gg 1$, which is
relevant for these astrophysical systems. Our present work instead focuses on a
different parameter regime of $\ell_{B} \sim 1$ and discusses the
self-regulation of local magnetization through synchrotron pair production;
consequently, it applies to a different set of astrophysical scenarios, as
discussed in Section~\ref{sec:application}.

\citet{2019ApJ...870...49S}, similar to \citet{2019ApJ...877...53H}, used
radiative QED-PIC simulations to study magnetic reconnection in the presence of
strong synchrotron cooling and copious $e^{\pm}$ pair production in the context
of pulsar and magnetar magnetospheres. However, the pair production mechanism
considered in their paper was photon interaction with a very strong
(approaching~$B_Q$) magnetic field, resulting in QED one-photon pair creation,
which has a completely different pair-production cross-section and rate. In that
regime, synchrotron emission also approaches the quantum limit where
$\epsilon_\mathrm{ph} \sim \gamma m_{e}c^{2}$, which alters the photon spectrum.

\citet{2021MNRAS.508.4532M} also studied radiative magnetic reconnection
self-regulated by pair production using analytic methods, but their main
interest was in $e^{\pm}$ pairs produced by inverse-Compton photons emitted by
reconnection-accelerated electrons upscattering an ambient photon field. They
argued that, since the IC radiation and the subsequent pair production are from
the same target mono-energetic photon field in the case under consideration, the
IC scattering needs to be in the Klein-Nishina regime to be able to produce
pairs, which affects the pair-production efficiency in these systems. Our study
is similar in spirit, but considers a different radiation mechanism and thus
applies to a different set of astrophysical systems.

Very recently \citet{2022arXiv220801882K} performed a calculation of the
synchrotron pair production multiplicity during a magnetic reconnection event,
which is similar to our present work, and applied it to low-luminosity AGN
including M87 and Sgr~A*. They found that for M87 parameters, once synchrotron
pair production kicks in, it lowers the upstream magnetization to
$\sigma\sim 8.7\times 10^{4}$, similar to our estimates. However, they called
this the ``low-energy flaring state'' and concluded that this state will not
produce MeV photons efficiently, and that $\sigma$ will grow again once the
pairs are advected from the region. The reason for their conclusion was that
they assumed reconnection only accelerates particles up to $\gamma \sim \sigma$,
which will not allow synchrotron photons to pair-produce in this low $\sigma$
state. As we have shown in the present paper, even an exponential cutoff above
$\gamma_\mathrm{max} \sim 4\sigma$ can allow the system to sustain a substantial
level of pair production in this low energy flaring state, making it difficult
for the system to spontaneously go back to a state with high magnetization.

\section{Discussion}
\label{sec:discussion}

We have considered synchrotron pair production in highly relativistic magnetic
reconnection, and studied how the $e^{\pm}$ pairs feed back on the reconnection
process itself, altering the magnetization close to the current sheet, and
ultimately reaching a self-regulated stable equilibrium~$\sigma_{\pm}$. We found
that this equilibrium is typically close to the threshold Lorentz factor for
synchrotron pair production, $4\sigma_\pm \sim \gamma_c$
[Equation~\eqref{eq:gamma_c}]. In other words, the equilibrium magnetization is
almost entirely determined by the local magnetic field strength. This process
provides an estimate for the magnetization in finite-$\ell_B$ astrophysical
systems where this quantity is poorly constrained, and we provided two examples
of such systems: M87 and the Crab pulsar. In both cases, synchrotron pair
production can reduce the initially very high magnetization to a much lower
level, significantly constraining the power that can go into VHE gamma-ray
emission.

Our model has adopted the assumption that relativistic magnetic reconnection
impulsively accelerates particles to a hard power-law spectrum with index
$p \sim 1$ and an exponential high-energy cutoff near
$\gamma_\mathrm{max} \sim 4\sigma$. This coefficient of 4 was originally
reported by \citet{2016ApJ...816L...8W} and remains uncertain up to a factor of
a few. The numerical value of this coefficient, however, turns out to be not too
important in our model, since pair production tends to push
$\gamma_\mathrm{max}$ close to~$\gamma_{c}$. Changing the numerical coefficient
to, e.g.,\ $\gamma_\mathrm{max}\sim 10\sigma$ only reduces the predicted final
magnetization~$\sigma_{\pm}$, but does not change the maximum extent of the
particle/radiation power law. The results in Section~\ref{sec:application} are
mostly independent of the exact energy of the exponential cutoff.

\citet{2021ApJ...922..261Z} recently demonstrated that in 3D reconnection with
moderate magnetization $\sigma = 10$, a secondary power law of $p\sim 1.5$
formed by free particles not captured by plasmoids can extend beyond
$\gamma_\mathrm{max}\sim 4\sigma$, up to a cutoff energy that scales linearly
with system size. Approximately ${\sim}20\%$ of the dissipated magnetic energy
goes into this secondary power law. This effect may potentially change the
conclusions of our model. Qualitatively, it will allow more pair production
activity at low~$\sigma$, effectively pushing the equilibrium magnetization even
lower. The quantitative effect of this secondary power law, especially at very
high magnetization $\sigma_{0}\gg 10$, will be studied in a future work.

In our model, we have neglected the change of the particle distribution, and hence of
the synchrotron spectrum, caused by copious pair production. In particular,
we have neglected the photons emitted by the secondary pairs. These photons in
general will have lower energies, and may serve as target photons for the much
higher-energy synchrotron photons close to the cutoff. However, near the final
equilibrium~$\sigma_\pm$, the power-law cutoff energy of synchrotron photons is
already close to the pair production threshold. Further increase of low-energy
photon density will only enable photons in the exponential tail to create pairs,
which will not meaningfully change the equilibrium magnetization. We have also
neglected synchrotron self-Compton (SSC) photons that may pair-produce. However,
since all synchrotron photons ultimately come from the dissipation of upstream
magnetic energy, $U_\mathrm{ph} \lesssim \beta_\mathrm{rec}U_B$, one expects the SSC
radiation energy density to be subdominant compared to the synchrotron energy
density by a factor of~$\beta_\mathrm{rec}$. Furthermore, SSC photons typically
will have much higher energies than the synchrotron photons.  Therefore, the density
of their scatter targets should in general be lower in a hard power-law photon
distribution with $\alpha < 1$. As a result, we expect SSC photons to play a
subdominant role in regulating the pair production equilibrium in a reconnection
event.

PIC simulations that incorporate photon-photon pair production, similar to what
was done by \citet{2019ApJ...877...53H} but in the regime of $\ell_{B}\sim 1$,
will help verify the validity of the analytic model presented in the present
work. Such time-dependent simulations will also capture the whole process of
photon emission and pair production, and will be able to measure how the
upstream magnetization responds to it. The results may place a more quantitative
bound on the gamma-ray luminosity from the relativistic reconnecting current
sheet before it is overwhelmed by $e^\pm$ pairs, therefore providing more
detailed estimates of the HE to VHE gamma-ray luminosity and spectra from these
systems.

\acknowledgments

We thank Yajie Yuan and Andrei Beloborodov for helpful discussions. AC and JD
acknowledges support from Fermi Guest Investigation grant 80NSSC21K2027. AC also
acknowledges support from NSF grant DMS-2235457. This work was also supported by
NSF grants AST-1806084 and AST-1903335, and by NASA grants 80NSSC20K0545 and
80NSSC22K0828. This work was stimulated by the discussions in the \emph{Fourth
  Purdue Workshop on Relativistic Plasma Astrophysics} in May~2022.

\bibliographystyle{aasjournal}

\end{document}